\documentclass{article}

    \PassOptionsToPackage{authoryear,round}{natbib}

\usepackage{natbib}

\usepackage[preprint]{neurips_2023}

\usepackage[utf8]{inputenc} 
\usepackage[T1]{fontenc}    
\usepackage{hyperref}       
\usepackage{url}            
\usepackage{booktabs}       
\usepackage{amsfonts}       
\usepackage{nicefrac}       
\usepackage{microtype}      
\usepackage{xcolor}         

\usepackage{amsmath,latexsym,amssymb,verbatim,amsbsy,times}
\usepackage{amsthm}
\usepackage{graphicx}
\usepackage{bbm}
\usepackage{bm}
\usepackage{caption}
\usepackage{algorithm}
\usepackage{algpseudocode}

\title{Constrained Max Drawdown: a Fast and Robust Portfolio Optimization Approach}

\author{
  Albert~Dorador \\
  Department of Statistics\\
  Uinversity of Wisconsin - Madison\\
  \texttt{albert.dorador@wisc.edu} \\
}

\newcommand{\R}{\ensuremath{\mathbb{R}}}   

\def \cN {\mathcal{N}}

\begin{document}

\maketitle

\begin{abstract}
We propose an alternative linearization to the classical Markowitz quadratic portfolio optimization model, based on maximum drawdown.  This model, which minimizes maximum portfolio drawdown,  is particularly appealing during times of financial distress, like during the COVID-19 pandemic.  In addition, we will present a Mixed-Integer Linear Programming variation of our new model that, based on our out-of-sample results and sensitivity analysis,  delivers a more profitable and robust solution with a 200 times faster solving time compared to the standard Markowitz quadratic formulation.
\end{abstract}

\section{Introduction}
Portfolio optimization is the process of finding the best possible portfolio of financial assets i.e. the best way to allocate a budget among different possible such assets, according to some objective. The objective typically maximizes factors such as expected return, and minimizes costs like volatility (often informally referred to as `financial risk').

The above optimization problem (maximizing expected return with minimum risk) is the backbone of Modern portfolio theory, which was introduced by  \cite{Markowitz1952} and further elaborated later by \cite{Markowitz1959}, earning the author the Nobel Prize in Economics in 1990.

The world's fascination by Markowitz ideas still continues today and many different variations of his model have been proposed over the last few decades trying to correct some of its drawbacks.

In a recent paper \citep{Dai2019}, the authors study one of the main drawbacks that the Markowitz model has: the sensitivity of optimal solutions to changes in the model parameters. In this project we will study this issue as well, and we will propose a simple way to improve the robustness of our solution.

Another aspect that used to be a significant drawback of the original Markowitz model was the quadratic nature of the model: a quadratic program may be considerably harder to solve than a Linear one, and a few decades ago the computational constraint was a factor that limited the model's applicability in practice. \cite{Konno1991} study this issue and propose a way to linearize the problem while providing a comparable performance. Hence, at least since the 1990's, portfolio optimization is no longer (exclusively) a quadratic optimization problem, but linear. 

In the present work, we propose an alternative linearization to the classical Markowitz quadratic portfolio optimization model, based on maximum drawdown, which we call `Max Drawdown' (MD).  This model, which minimizes maximum portfolio drawdown,  is particularly appealing during times of financial distress, like during the COVID-19 pandemic. 

The rest of the paper is organized as follows: section 2 reviews both the Markowitz model (along with two simple variations) as well as the mean absolute deviation model \citep{Konno1991}, and it introduces the MD model. Section 3 implements the models introduced in section 2 (minus \cite{Konno1991}) and presents in-sample and out-of-sample results.  In section 4 we analyze the results presented in the previous section, including a sensitivity analysis.   Section 5 concludes.

\section{Mathematical models}

In all deterministic portfolio optimization models the general assumption is that parameter values are correctly observed (i.e. sample values correspond to true population values) and are fixed, at least to a relevant extent and for sufficiently long. These assumptions are of course naive. However, improving on these assumptions would require much more advanced tools. Moreover,  it is not clear how much it would be gained by doing that since, at the core, portfolio optimization is about choosing the assets that will statistically perform best jointly, but this inevitably has a component of `forecasting the future' which, in finance, is an impossible task in abstract, and some assumptions must be made (e.g. the distribution of returns stays constant during the investment horizon, or at most it changes according to a known model).

Having said that, there is certainly room for improvement on this front, and some relaxation of the assumptions, particularly in the form of using stochastic programming, is likely to lead to an improvement in out-of-sample performance.

\subsection{Model 1: Classical Markowitz model}

Below we review the Nobel prize - winning Markowitz model (see \cite{Markowitz1952} and \cite{Markowitz1959}).
$$\begin{aligned}
  \underset{x \in \R^n}{\text{min}}\qquad& \sum_{i=1}^n \sum_{j=1}^n \sigma_{i,j} x_i x_j\\
    \text{subject to:}\qquad& \sum_{i=1}^n r_i x_i \geq \rho && (1)\\
    & \sum_{i=1}^n x_i = 1 && (2)\\
    & 0 \leq x_i \leq 1 && i=1,\dots,n && (3)
    \end{aligned}$$

Where the decision variable $x_i$ is the proportion of the funds allocated to asset $i$, $\sigma_{i,j}$ is the (daily) covariance between asset $i$ and asset $j$, $r_i$ is the expected return of asset $i$ (computed via the Method of Moments i.e. the arithmetic mean over the sample period), and $\rho$ is a minimum (expected) return that is required per trading day.

The objective function seeks to minimize the standard deviation of the portfolio by considering the variance of each asset as well as its covariance with respect to the other assets in the sample.

Constraint (1) ensures that the optimal asset allocation achieves a minimum acceptable expected return.

Constraint (2) enforces that exactly 100\% of the funds are allocated to some combination of assets.

Constraint (3) is a set of $n$ box constraints that ensures each asset is assigned a non-negative proportion of the funds, up to a maximum of 100\%.

Therefore, this model has a quadratic objective function (since the decision variables are squared) and a set of linear constraints, so it is a QP.

Since the covariance matrix is, by definition, positive semidefinite, then: this is a convex quadratic program, the feasible set is a polyhedron, the solution lies either on the boundary or the interior of the feasible set, and it is relatively easy to solve (compared to a quadratic program with a non positive definite parameter matrix in the objective function).

In that direction, it is worth pointing out that Slater's conditions are satisfied:

Let $x^T\Sigma x$ be the matrix form of the objective, where $\Sigma$ is the covariance matrix.

Note that the program is indeed convex: $x^T\Sigma x$ is convex since $\Sigma$ is positive semi-definite, and all constraints are affine, and therefore convex (and concave too)

Further note the program is strictly feasible: let $x_i = 1$, for a given stock $i$ with an expected return strictly above $\rho$ (there must exist at least 1 such stock, otherwise the program is unfeasible); then the inequality constraint is satisfied strictly, and the equality constraint is also satisfied

Therefore, strong duality holds and, since we know a finite $p^*$ exists (since this is a convex quadratic program), then we are guaranteed a finite $d^*$ exists too and $p^* = d^*$ (where $p^*$ and $d^*$ are the optimal objective values of the Primal and the Dual, respectively).

In our implementation this model has about 400 continuous variables and 400 constraints (including box constraints).

\subsection{Model 2: Reverse Markowitz model}

This is the most basic variation of the classical Markowitz model, which is not as popular as the original Markowitz,  neither in academia nor industry. One possible reason is that it is much more arbitrary to set a maximum standard deviation than a minimum return (for which you have clear minimum acceptable thresholds like bond yields of maturities similar to one's sample horizon). Furthermore,  like in our particular case, the data collected does not allow us to require a very low daily portfolio standard deviation (otherwise the optimization problem is unfeasible i.e. the feasible region is empty).

$$\begin{aligned}
  \underset{x \in \R^n}{\text{max}}\qquad& \sum_{i=1}^n r_i x_i \\
    \text{subject to:}\qquad& \sum_{i=1}^n \sum_{j=1}^n \sigma_{i,j} x_i x_j \leq \sigma_0 && (1)\\
    & \sum_{i=1}^n x_i = 1 && (2)\\
    & 0 \leq x_i \leq 1 && i=1,\dots,n && (3)
    \end{aligned}$$
    
Where the decision variable $x_i$ is the proportion of the funds allocated to asset $i$, $\sigma_{i,j}$ is the covariance between asset $i$ and asset $j$, $r_i$ is the expected daily return of asset $i$ (computed via the Method of Moments i.e. the arithmetic mean over the sample period), and $\sigma_0$ is a maximum acceptable daily standard deviation over the time horizon of the sample period.

The objective function seeks to maximize the expected return of the portfolio.

Constraint (1) ensures that the standard deviation of the optimal asset allocation does not go beyond a given threshold.

Constraint (2) enforces that exactly 100\% of the funds are allocated to some combination of assets.

Constraint (3) is a set of $n$ box constraints that ensures each asset is assigned a non-negative proportion of the funds, up to a maximum of 100\%.

Therefore, this model has a linear objective function and a set of quadratic and linear constraints. Note it cannot be considered a linear program because even if the objective is linear, not all constraints are. This is in fact a Quadratically Constrained Quadratic Program (QCQP).

In our implementation this model has, again,  about 400 continuous variables and 400 constraints (including box constraints).

\subsection{Model 3: Simultaneous mean-variance optimization}

In both Model 1 and Model 2 one's preferences are heavily constrained by the random data one collects, in the sense that one cannot require a minimum return higher than any return observed in the sample collected, and likewise one cannot require a maximum variance lower than any variance observed in the sample collected.

By optimizing both the expected return and the standard deviation simultaneously we overcome the issue above, in the sense that we will not have an unfeasible problem due to 'unrealistic market demands'.

However, this comes at the cost of introducing arbitrariness in the form of the penalty parameter $\lambda$, which balances how much we care about each of the two objectives.

In any case, as we will see in a later section, we will present a simple heuristic that attempts to remove part of this arbitrariness.

$$\begin{aligned}
  \underset{x \in \R^n}{\text{min}}\qquad& -\sum_{i=1}^n r_i x_i + \lambda\sum_{i=1}^n \sum_{j=1}^n \sigma_{i,j} x_i x_j\\
    \text{subject to:}\qquad& \sum_{i=1}^n x_i = 1 && (1)\\
    & 0 \leq x_i \leq 1 && i=1,\dots,n && (2)
    \end{aligned}$$

Where the decision variable $x_i$ is the proportion of the funds allocated to asset $i$, $\sigma_{i,j}$ is the (daily) covariance between asset $i$ and asset $j$, $r_i$ is the expected daily return of asset $i$ (computed via the Method of Moments i.e. the arithmetic mean over the sample period).

The objective function seeks to minimize the negative of the expected return (i.e. equivalent to maximizing the expected return) while also minimizing $\lambda$ times the variance of the portfolio.

Constraint (1) enforces that exactly 100\% of the funds are allocated to some combination of assets.

Constraint (2) is a set of $n$ box constraints that ensures each asset is assigned a non-negative proportion of the funds, up to a maximum of 100\%.

Therefore, this model has a quadratic objective function and a set of linear constraints, so it is a QP.

Finally, it is worth mentioning that a variation of the above three models introducing $L_1$ regularization was considered. However, it was eventually discarded since, by construction, it had no effect irrespective of the size of the penalty parameter $\mu > 0$. We provide a simple proof next.

$L_1$ regularization introduces an extra term in the objective function, which, in case of a minimization problem, is $\mu \sum_{i=1}^n u_i$, with $|x_i| \leq u_i$ for all $i \in \{1,\dots,n\}$. But then, since $x_i \geq 0$ for all $i \in \{1,\dots,n\}$, the previous epigraph condition simplifies to $x_i \leq u_i$ for all $i \in \{1,\dots,n\}$, which implies $\sum_{i=1}^n x_i \leq \sum_{i=1}^n u_i$. But we require $\sum_{i=1}^n x_i = 1$ so $1 \leq \sum_{i=1}^n u_i$. Finally, since $\sum_{i=1}^n u_i$ is penalized in the objective by $\mu > 0$, and there are no additional constraints on $u_i$, we then have $\sum_{i=1}^n u_i = 1$ at the optimum, irrespective of the particular (positive) value of $\mu$.

In our implementation this model has, again, about 400 continuous variables and 400 constraints (including box constraints).

\subsection{Model 4: Mean Absolute Deviation (Konno and Yamazaki 1991)}

The authors propose in their 1991 paper [4] a way to turn portfolio optimization into a linear program based on the so-called `epigraph trick'. Their key idea is to substitute the standard deviation for the mean absolute deviation, which is a very related concept, as we show next:

(Sample) Mean absolute deviation: $\frac{1}{T} \sum_{t=1}^T \left| \sum_{i=1}^n (r_{i_t} - \overline{r_i})x_i \right|$ 

(Sample) Standard deviation: $\sqrt{\frac{1}{T} \sum_{t=1}^T [\sum_{i=1}^n (r_{i_t} - \overline{r_i})x_i]^2}$

where $\overline{r_i} = \sum_{t=1}^T x_i$

The above two expressions would actually be equal if the square root was placed in the innermost sum (recall that $\sqrt{f(x)^2} = |f(x)|$). In fact, Konno and Yamazaki prove that the two resulting optimization problems are equivalent if the returns are multivariate normally distributed.

Original program (non linear):

$$\begin{aligned}
  \underset{x \in \R^n}{\text{min}}\qquad& \frac{1}{T} \sum_{t=1}^T \left| \sum_{i=1}^n (r_{i_t} - \overline{r_i})x_i \right|\\
    \text{subject to:}\qquad& \sum_{i=1}^n r_i x_i \geq \rho && (1)\\
    & \sum_{i=1}^n x_i = 1 && (2)\\
    & 0 \leq x_i \leq 1 && i=1,\dots,n && (3)
    \end{aligned}$$

Linear program using epigraph trick:

$$\begin{aligned}
  \underset{x \in \R^n, y \in \mathbb{R^T}}{\text{min}}\qquad& \frac{1}{n} \sum_{t=1}^T y_t\\
    \text{subject to:}\qquad& y_t \geq  \sum_{i=1}^n (r_{i_t} - \overline{r_i})x_i && t=1,\dots,T && (1)\\
    & y_t \geq  -\sum_{i=1}^n (r_{i_t} - \overline{r_i})x_i && t=1,\dots,T && (2)\\
    & \sum_{i=1}^n r_i x_i \geq \rho && (3)\\
    & \sum_{i=1}^n x_i = 1 && (4)\\
    & 0 \leq x_i \leq 1 && i=1,\dots,n && (5)
    \end{aligned}$$
    
We see the epigraph variable $y_t$ replacing the mean absolute deviation in the objective function, as well as the two sets of $T$ epigraph constraints. The rest of the constraints are identical to Models 1, 2 and 3.

As the authors discuss, this linearized model has several advantages over the usual quadratic one, and here we highlight the ones we believe are most important, in order of importance:

First, the number of functional constraints remains constant at $2T + 2$ regardless of the number of stocks included in the model, so it allows one to update the optimal portfolio in real time even if the universe of stocks under consideration is thousands of stocks (which is a very standard number in modern capital markets).

Second, an optimal solution contains at most $2T + 2$ non-zero components if short-selling is allowed (non-negativity constraint is dropped). In contrast, an optimal portfolio derived from $L_2$ risk (instead of $L_1$ like in this case), may contain as many as $n$ assets with non-zero allocation. This is in general not desirable due to fixed costs (transaction fees) and increased portfolio management difficulty (more companies and exposures to monitor). Note that controlling for how many stocks have non-zero allocations via a lower bound on the decision variable (that is activated only if the stock is selected) would turn the problem into a Mixed Integer Program, which is in theory even harder to solve than a QP. We will explore this option in Model 6 and we'll see in practice this is not the case.

Third, there's no need anymore to calculate a covariance matrix, so it is easier and faster to update ones' optimal portfolio when new stocks are added into the universe under consideration, or when new data on individual stock returns becomes available. However, related to the second point above, this model may favor stock concentration (i.e. the opposite of stock diversification) since it does not use covariance information. This may not be desirable.

Overall, the authors report that the computational complexity of solving a $L_1$-based model is $O(n)$ while that of a traditional $L_2$-based model is $O(n^2)$, where $n$ is the number of assets under consideration, and so the time savings for large $n$ can be very significant. Or, for a fixed solving time, the LP can consider a much larger universe of stocks, providing a much better optimal solution (at least in theory). As a side note, I believe their linear-time claim is based on empirical evidence since, for example, the complexity of a very common LP algorithm like the Simplex method is exponential in the worst case, but it has been observed to perform linearly in most cases.

But, in addition to the above advantages, we would like to add two key advantages that an LP has over a QP in any setting, not just in the case of portfolio optimization.

First, the optimal solution of an LP, if it exists, is always at a vertex of the polyhedral feasible region, so the optimum of any LP can always be found by checking a finite number of points (which is the essence of the Simplex method). On the other hand, in a QP the optimum may be reached in the interior of the feasible region, and hence there is absolutely no guarantee that it can be found by checking a finite number of points, and hence other, more sophisticated methods must be used. 

Second, the objective function of an LP is by definition always convex (and concave), whereas this is not always the case in a QP (the coefficient matrix may not be positive semi-definite), in which case the problem becomes much harder to solve. In portfolio optimization, though, this second aspect will never be an issue since the covariance matrix is by definition positive semi-definite. But still, it is worth pointing this general advantage of LP compared to QP.

If we were to implement this model, it would have about 450 continuous variables and 520 constraints (including box constraints).

Next we present our proposed linearization of the portfolio optimization problem.

\subsection{Model 5: Maximum Drawdown portfolio optimization}

We propose a new way to optimize a portfolio that relies on linear programming. The idea is simple: find a portfolio of stocks that minimizes the maximum drawdown of the portfolio (subject to a minimum expected return). \cite{Chekhlov2004} consider a similar approach but they opt for maximizing expected return subject to a maximum drawdown allowed. We instead opt for the opposite formulation, which we find perhaps more intuitive (similar to how the classical Markowitz framework might be considered more intuitive than the `reverse' Markowitz).

The `maximum drawdown' of an asset over a time period is defined as the largest drop in price of the asset in any given period. In our case, the optimal portfolio will combine stocks such that the worst overall performance of the portfolio in any given trading day is as good (`less bad') as possible.

This is a reasonable goal to pursue in building a portfolio, and it has several advantages over the traditional mean-variance portfolio optimization. Most notably, the fact that we consider a downside risk metric instead of a general risk metric like the standard deviation, which also penalizes desirable outcomes like presence of returns above the mean return.

Hence, we wish to maximize the minimum portfolio return over a sample time period (equivalent to minimizing the maximum portfolio negative return).

We use the epigraph trick as means of linearization. To prevent a trivial solution allocating 100\% of the funds to the one stock with the minimum maximum drawdown (largest minimum return in any given trading day of the sample period), we impose a ceiling of 50\% of the funds on any given allocation, so there will be at least 2 stocks in the optimal portfolio (and note that it is not necessarily the case that they will be the 2 stocks with largest minimum return, since those may happen on the same day and hence add up to a larger maximum portfolio drawdown than the optimal achievable). Note that by setting the ceiling to 50\% we avoid a trivial solution while also avoiding the arbitrariness of imposing a minimum number of $k > 1$ stocks in our optimal portfolio. Finally, note that we don't really lose anything by requiring at least 2 stocks in the optimal portfolio, since basic domain knowledge tells us that some degree of diversification is always good, and with 1 stock you have none.

Original program (non-linear):

Define $\overline{r_t} = \sum_{i=1}^n r_{i_t} x_i$ , that is, the weighted average return on day $t$ of a portfolio using weights $x$.

$$\begin{aligned}
  \underset{x \in \R^n \, \, t=1,\dots,T}{\text{max min}}\qquad& \overline{r_t}\\
    \text{subject to:}\qquad& \sum_{i=1}^n r_i x_i \geq \rho && (1)\\
    & \sum_{i=1}^n x_i = 1 && (2)\\
    & 0 \leq x_i \leq 0.5 && i=1,\dots,n && (3)
    \end{aligned}$$

Linear program using epigraph trick:

$$\begin{aligned}
  \underset{x \in \R^n, y \in \mathbb{R}}{\text{max}}\qquad& y\\
    \text{subject to:}\qquad& y \leq \overline{r_t} && t=1,\dots,T && (1)\\
    & \sum_{i=1}^n r_i x_i \geq \rho && (2)\\
    & \sum_{i=1}^n x_i = 1 && (3)\\
    & 0 \leq x_i \leq 0.5 && i=1,\dots,n && (4)
    \end{aligned}$$

The epigraph trick can be interpreted as follows: find the largest possible value of the single variable $y$ such that, for each and every day, $y$ is below (or equal to) the portfolio return given by the weights vector $x$. Therefore this achieves the goal of finding the budget allocation $x$ that maximizes the minimum portfolio return on any given day. The rest of constraints are the same as in the previous models.

Note our LP has just $T+2$ constraints, so it's even simpler than Model 4. But even more importantly, the number of constraints is independent of the number of stocks under consideration, so it scales very well to larger sets of stocks.

In addition, our LP has $n+1$ variables, instead of $n+T$ in Model 4, and just 1 variable more than the standard QP.

Finally, we write this LP in standard form. By writing it in standard form, this program does not really gain anything (in fact, it becomes somewhat harder to read) but it may be interesting to show one example of an LP that uses the epigraph trick and is in standard form.

$$\begin{aligned}
  \underset{x \in \R^n, y \in \mathbb{R}}{\text{max}}\qquad& y\\
    \text{subject to:}\qquad& y - \overline{r_t}\leq 0 && t=1,\dots,T && (1)\\
    -& \sum_{i=1}^n r_i x_i \leq -\rho && (2)\\
    & \sum_{i=1}^n x_i \leq 1 && (3.1)\\
    -& \sum_{i=1}^n x_i \leq -1 && (3.2)\\
    & 0 \leq x_i \leq 0.5 && i=1,\dots,n && (4)
    \end{aligned}$$

In our implementation this model has about 400 continuous variables and 450 constraints (including box constraints).

\subsection{Model 6: Maximum Drawdown portfolio optimization with minimum allocation constraint}

In this variation of Model 5 we impose a minimum allocation size, which in turn implies a maximum number of positive allocations. This is a potentially more refined formulation since having many small allocations is not desirable in practice, as we have already discussed in subsection 2.4. However, this variation is, in theory, much harder to solve, so it may not be worth it depending on the problem structure, the number of stocks ($n$) and sample time periods ($T$) under consideration.

Let $M = 0.5$ be a `big M' parameter. Note $0.5$ is the smallest (i.e. 'best') $M$ can be because $x_i$ could be as large as 0.5 (but not more, since 0.5 is the (least) upper bound on any allocation by construction). In fact, by choosing $M$ this way we obtain the tightest LP relaxation possible; this way the feasible region we work with is as close as possible to the convex hull using available information from the problem.

We want to model for all $i \in \{1,\dots,n\}$ $x_i > 0 \Rightarrow x_i \geq 5\%$. This way we also ensure that at most 20 stocks (out of about 400) will be picked.

Let $z_i$ $i=1,\dots,n$ be a binary variable that is 1 if and only if $x_i > 0$.

Note that if we impose $x_i \geq 0.05z_i$ then when $z_i = 1$ indeed $x_i \geq 0.05$, and when $z_i = 0$ then $x_i \geq 0$... but the problem is that if we don't do anything else the solver will dodge this (attempted) logical constraint by just setting $z_i = 0$ for all $i$. We can fix this by also requiring $x_i \leq M z_i$ where $M = 0.5$ is a 'big M' parameter. With this additional constraint, the solver cannot dodge our previous constraint by setting all $z_i = 0$ because that would imply $0 \leq x_i \leq 0$ for all $i$, which is not even a feasible solution since we require $\sum_{i=1}^n x_i = 1$.

MILP using epigraph trick and logical constraints:

$$\begin{aligned}
  \underset{x \in \R^n, y \in \mathbb{R}, z \in \{0,1\}^n }{\text{max}}\qquad& y\\
    \text{subject to:}\qquad& y \leq \overline{r_t} && t=1,\dots,T && (1)\\
    & \sum_{i=1}^n r_i x_i \geq \rho && (2)\\
    & \sum_{i=1}^n x_i = 1 && (3)\\
    & x_i \geq 0.05z_i && i=1,\dots,n && (4)\\
    & x_i \leq M z_i && i=1,\dots,n && (5)\\
    & 0 \leq x_i \leq 0.5 && i=1,\dots,n && (6)
    \end{aligned}$$
    
As it can be seen, implementing a very minor refinement like the one in Model 6 required turning an LP into a MILP and increased the number of variables from $n+1$ to $2n+1$, and the number of constraints from $T+2$ to $T+2+2n$, which means the number of functional constraints is no longer independent of the number of stocks.

But more fundamentally, this problem seems similar to the famous Subset sum problem (a special case of the even more famous Knapsack problem \citep{Mathews1896}), consisting of finding a subset of numbers in a set that sum to a given constant, in this case 1,  which is NP-complete. Our problem at hand is at least as hard, since in our case the candidates to add up are not given ex-ante, which means that there are $2^n$ total possible combinations and checking all of them by brute force would would take thousands of times the current age of the universe for $n=100$. In our case we have $n=393$. However, as we will see, the particular structure of our problem and the use of a sophisticated MIP solver like Gurobi will in fact produce a solution in less time than our LP in Model 5.

In our implementation this model has about 400 continuous variables, 400 binary variables, and 1250 constraints (including box constraints).

\section{Data and model deployment}

Since our proposed MD models minimize the maximum portfolio loss in any given trading day, it stands to reason that it may be particularly useful in times of financial distress.

Therefore, we use early COVID-19-era market data on about 400 US stocks from the S\& P 500 index,  publicly available on Yahoo Finance and automatically downloaded through an in-house Python script. Average return vector and covariance matrix are then computed from these data. 

We consider a 50-50 train/test split for our data. Concretely, we use February 1 -- May 1, 2020 data to train and May 2 -- August 1, 2020 to test.

As mentioned earlier, we need a heuristic to choose the tuning parameter $\lambda$. Cross-validation is tricky in a time-series context. Instead, we consider the following approach.

We observe that in our universe of optimal portfolios for 100 different values of $\lambda$ (equally spaced from $10^{-3}$ to $10^4$),  the minimum portfolio standard
deviation is about 0.5930\% daily, and the maximum expected portfolio daily return is 1.3398\%. We want to find the value of the parameter $\lambda$ that yields an optimal portfolio with minimum Euclidean norm from the point (0.5930,1.3398).  The in-sample optimal value of $\lambda$ is found to be approximately 0.08.

The rationale behind this heuristic is that the top left corner is the ideal situation (high expected return and low risk), and in our universe of optimal portfolios for different values of the top left corner is given by the two coordinates we have just mentioned. This is, hence,  a natural heuristic to objectively choose the (in-sample) `best' Pareto optimal portfolio, i.e. a way to objectively choose the `best' balance between the goal of maximizing expected return and minimizing standard deviation.  As a reminder, a Pareto optimal portfolio is a portfolio in which it is not possible to increase the portfolio expected return without also increasing the portfolio standard deviation, and vice versa (it is not possible to reduce the portfolio standard deviation without also reducing the portfolio expected return). Any portfolio to the
left of the Pareto curve (also called `Efficient frontier' in this context) is unfeasible, and any portfolio to the right of the Pareto curve is inefficient i.e. it is possible to improve one of the two metrics without worsening the other.

\subsection{In-sample results}

Table 1 below shows the in-sample results from solving the portfolio optimization models that we have considered in the present work.  The quantities in the first three columns below are daily figures.

\begin{table}[htbp]
\caption{In-sample performance of  portfolio optimization models surveyed}
\begin{center}
\begin{tabular}{lccccc} 
 \hline
 Model & Exp. Return & Std. Deviation & Max Drawdown & \# Stocks & Time (s) \\
\hline
 Markowitz & 0.18\% & 0.6\% & -2.0\% & 16 & 0.02\\ 
 Markowitz (reverse) & 0.59\% & 1.0\% & -3.2\% & 13 & 0.02\\ 
 Simultaneous optimization & 0.70\% & 1.4\% &-4.8\% & 12 & 0.02\\ 
 MD & 0.22\% & 0.8\% & -0.9\% & 9 & 0.003\\ 
 MD (constrained) & 0.23\% & 0.8\% & -0.9\% & 5 & 0.0001\\ 
 \hline
\end{tabular}
\end{center}
\end{table}

\subsection{Out-of-sample results}

Table 2 below shows the out-of-sample analogs of the results in Table 1.

\begin{table}[H]
\caption{Out-of-sample performance of  portfolio optimization models surveyed}
\begin{center}
\begin{tabular}{lcccc} 
 \hline
 Model & 3-Month Return & Daily Return & Std. Deviation & Max Drawdown\\
\hline
 Markowitz & 5.7\% & 0.09\% & 1.7\% & -6.1\%\\ 
 Markowitz (reverse) & 5.3\% & 0.08\% & 1.7\% & -7.2\% \\ 
 Simultaneous optimization & 5.2\% & 0.08\% & 2.1\% &-8.6\%\\ 
 MD & 9.4\% & 0.15\% & 2.3\% & -3.3\% \\ 
 MD (constrained) & 8.5\% & 0.13\% & 2.0\% & -3.3\%\\ 
 \hline
\end{tabular}
\end{center}
\end{table}

\section{Results analysis}

\subsection{In-sample solution comparison}

The in-sample results show that the price to pay (in terms of expected return or standard deviation) in order to minimize the maximum drawdown is arguably not very high: those models with higher expected return also have higher standard deviation (and vice versa), while always having much larger maximum drawdown (as it's not optimized for).

In addition, the MD models are much faster. Concretely, our MILP variation is the fastest, by far: 200 times faster than the QPs, and 25 times faster than the unconstrained MD (LP). This is due to the fact that Gurobi is a highly optimized solver for MIP and is able to take advantage of the the particular structure of our portfolio optimization problem. According to Gurobi, their sophisticated MIP algorithm (a combination of Branch-and-Bound and Cutting plane method, i.e. the fancier Branch-and-cut) first does some heavy preprocessing to reduce the problem dimension, and then solves a linear programming relaxation of the problem via Simplex iterations; it seems plausible that by requiring a minimum of a 5\% allocation if a stock is chosen, this might already discard many `bad' stocks based on the objective in the preprocessing stage, which greatly reduces the combinatorial complexity of the MILP. Further investigation suggests that after adding just 2 cuts to the original LP relaxation and solving that instance, Gurobi managed to recover an integer feasible solution (and hence solving the entire original MILP) without even needing to do any branching. 

\subsection{Out-of-sample solution comparison}

In this subsection we analyze the results obtained by each model trained with data from February 1 to May 1 of 2020 (three months), and tested on the subsequent three months (May 2 - August 1). This way we can compare the portfolio optimization solution out of sample. 

The idea is to simulate what would have happened if we had optimized our portfolio according to each of the 4 models above on May 1 with data from the previous three months, and then we had checked on August 1 how well each portfolio did. The reason why we wait until August 1 to check is because we optimize our portfolio with 3-month data so it's only natural the investment horizon should also be 3 months (with the implicit and naive assumption that history repeats itself at least to some relevant extent).

As previously reported in Table 2, our proposed models yielded substantially higher out-of-sample returns than any other model surveyed. Of course, these results should be considered preliminary and hence taken with a grain of salt, since a similar experiment should be repeated multiple times to gather enough evidence to conclude which model is likely the `best'.

Unsurprisingly, we notice that the out-of-sample performance (Table 2) is significantly worse than the in-sample performance (Table 1), which cautions the reader against overly optimistic expectations when trying to use the past to gain an edge into the future.

\subsection{Sensitivity analysis}

\subsubsection{Right hand side (RHS) of primal constraints} 

In general, in all models considered, the only RHS that is in our opinion potentially interesting to check is the RHS of the constraint $\sum_{i=1}^n x_i = 1$, and it can be interpreted as the value of having access to 1 unit more of budget e.g. through borrowing (shadow price of this constraint). However, a more careful analysis reveals that, since we work with stock returns and not absolute monetary gains, it is in fact pointless to ask how valuable it would be to have an additional unit of budget. Here's why: stock returns are invariant to budget size (only the different proportions of the budget invested matter), and the standard deviation  of the portfolio in fact increases as the budget increases, which after more careful thought seems obvious because if you have more budget you can, well, lose more budget.

The same argument applies to Model 5, since the maximum drawdown is, at the end of the day, a return, and so invariant to budget size.

For Model 6 no such sensitivity analysis is possible since there's no well-defined concept of `dual' in MIP.

\subsubsection{Primal objective coefficients}

The primal objective coefficients of Models 4, 5 and 6 are not interesting since they are just auxiliary variables used in the epigraph trick.

On the other hand, it may be interesting to see how sensitive the solution (vector of budget allocations) is to changes in the vector of expected returns and/or (careful) changes in the covariance matrix, for now restricting attention to when they appear in the objective. Thus, these changes leave the original primal feasible region unchanged, so the original optimal solution is still feasible. For ease of exposition, let's focus on Model 2 (maximize expected return subject to a maximum allowable variance). In this new scenario, the dot product of the original optimal solution vector and the new vector of expected returns is a lower bound to the new maximization problem. To obtain an upper bound, though, we would need to find a feasible point of the (new) Dual (by Weak Duality in convex programs). However, finding the Dual is not trivial since the Primal is a quadratically constrained quadratic program and the Lagrangian is a function of $x, \lambda , \nu$ where $x$ is an n-dimensional vector, which suggests the use of more sophisticated methods (e.g. matrix differentiation). Even if that was not an issue, the resulting Dual objective is very messy, as the reader can verify below (in un-simplified form):

$$\frac{-r-\nu e}{2\lambda}\Sigma^{-1}r + \frac{-r-\nu e}{2} \left(\frac{-r-\nu e}{2\lambda}\right)^T - \lambda \sigma_0 + \nu \frac{-r-\nu e}{2\lambda}\Sigma^{-1} e -\nu$$

Note $e$ is an n-dimensional vector of ones, and $\Sigma$ is the covariance matrix. Also note $\Sigma^{-1}$ exists since the covariance matrix of risky assets is positive definite, so has nonzero (positive) determinant.

At this point we still need to find a feasible dual point. Note that even if we manage to somehow find one by inspection, the upper bound that it would provide to the new primal may be arbitrarily bad.

However, we can still obtain valuable insight by adding a small perturbation to the data on stock returns, which will then automatically add some perturbation to the covariance matrix (and will do so preserving its defining properties since we are not perturbing the covariance matrix directly 'by hand'). Then we can see how the optimal budget allocation of Models 1, 2, 5 and 6 change.

Note that if we go this route we are no longer restricted to making changes in the coefficients of the primal objective or primal RHS (dual objective coefficients), but also in the coefficients of the left hand side (LHS) of the primal constraints. 

We have performed this empirical sensitivity analysis in the last portion of the previous section and here we discuss the results.

We have modeled the random perturbation of the return of stock $s$ at time $t$, $p_{s,t}$ that we mentioned previously, as follows:
$$p_{s,t} = \cN(0,\sigma_s)/c$$
where $\cN(0,\sigma_s)/c$ denotes a Normal distribution centered at $0$ and with standard deviation $\sigma_s$, divided by a positive constant $c$, which in our case is $10^3$ to suit the scale of our daily returns.

Observe that the size of the random perturbation depends on the standard deviation of each stock, which is a more realistic model assumption compared to a fixed random shock applied to all stocks (or completely arbitrary perturbations) since more volatile stocks should have bigger price fluctuations across time.

We have decided to measure the change in covariance matrix by computing the average absolute pairwise difference between the original and perturbed covariance matrices. This is similar to the Frobenius norm of the difference matrix, but has a more straightforward interpretation: average absolute change in each element of the covariance matrix is $0.11755\%^2$. Considering that the average size of an element in the original covariance matrix is $4.2113\%^2$, the size of the perturbation represents roughly a 2.8\% change in each covariance element, on average.

There are several different ways to capture the change in allocation, and in the code section above the reader can glance at various metrics printed. However, we believe that the most meaningful metric is to see the percentage by which the new allocation differs from the original one, restricting the comparison to the positive original allocations (because if the original allocation is $0$ and now is positive, the change would be something like `$+\infty \%$' which is not very meaningful and would completely dilute all other finite changes). Then we take the average of the absolute value of those percentage differences.

Table 3 below shows that the average absolute allocation change in each model is between $38\%$ and $40\%$ for the Markowitz models, $47\%$ for the linear Maximum Drawdown model, and only $3.7\%$ for the MILP Maximum Drawdown model. With the exception of the last model, and considering that on average each covariance element was changed by 2.8\% only, the sensitivity analysis performed reveals the fragility of the optimal solution to a small change in the parameter values.

\begin{table}[H]
\caption{Average absolute change in allocation after random perturbation}
\begin{center}
\begin{tabular}{lc} 
 \hline
 Model & Avg abs change in allocation\\
\hline
 Markowitz & 38.1\% \\ 
 Markowitz (reverse) & 39.8\% \\ 
 Simultaneous optimization & 42.3\% \\ 
 MD & 47.1\% \\ 
 MD (constrained) & 3.7\% \\ 
 \hline
\end{tabular}
\end{center}
\end{table}

The difference in robustness is striking. The explanation is as follows: the 'pure' Maximum Drawdown model is inherintly less robust than a Markowitz model since it has an extreme metric (the minimum observed return) as the optimization objective. However, the Maximum Drawdown model becomes more robust once we disallow small allocations (below $5\%$ in our case), and so it is expected that at least the choice of (fewer) stocks is more similar to the original model, and so we have a better chance of avoiding $-100\%$ changes (stocks dropped). In general, each stock selected in the original model should be 'more strongly selected' than in the 'pure' version of the model, since it is selected knowing that at least we will allocate $5\%$. Intuitively, we are reducing overfitting, and hence increasing the robustness of the model.

Therefore, it seems that a simple way to increase the robustness of any given portfolio optimization model is to require a minimum allocation for each selected stock (at the expense of turning the portfolio optimization model into a MILP, but it seems this is not necessarily an issue for problems with a size and structure similar to ours, especially with today's solvers and computational power). Naturally, the previous claim would require a deeper analysis to be more conclusive, but it seems at least a reasonable conjecture in light of the evidence we have and the intuitive connection with overfitting.

There is no denying that our attempt at increasing the robustness of the optimal solutions by requiring a minimum allocation is less sophisticated than the one \cite{Dai2019} and other authors before them discuss, which is based on modeling parameter uncertainty, venturing into the realm of stochastic programming. This is certainly a possible extension to this work. Another way \cite{Dai2019} try to improve the robustness of the solution is by promoting sparsity through $L_1$ regularization. This is interesting, because we have proved earlier that $L_1$ regularization has no effect in portfolio optimization. The answer to this apparent contradiction is that\cite{Dai2019} allow short-selling, i.e. they do not impose $x_i \geq 0$ for all $i$, while we do, and we used this fact quite crucially in our proof.

\subsection{Analysis of Model 6's (constrained MD) optimal solution}

Due to its improved robustness while providing comparable risk-adjusted performance, Model 6 (MILP formulation of the MD model) is our model of choice, at least for a portfolio optimization problem of a similar size to this one (about 400 stocks). Thus, we will focus our solution analysis on the optimal solution this model has found.

According to our proposed model, the three stocks with largest allocation should be CLX (40.34\%), KR (28.92\%), and TIF (20.73\%). But what do those tickers mean? The first one will make a lot of sense to the reader: Clorox. Clorox is a major producer of cleaning products, especially those containing bleach and other strong disinfectants. Needless to say, it seems reasonable that such a stock is picked during a pandemic. Models 1 (12.77\%) and 2 (4.8\%) also pick this stock, but with less emphasis. 

The second one is Kroger, the largest supermarket in the US by revenue. There are three main reasons why, in the time-frame of our training data (February-May 2020) it makes sense to have a significant allocation on this stock. First, with restaurants being closed or operating at reduced capacity due to the pandemic, supermarkets have seen a rise in demand. Second, Kroger in particular has been able to benefit more than other supermarkets thanks to their investments made in their online sales service since 2017, which allowed Kroger to absorbe the surge in demand, with digital sales climbing 92\% in the three months before May 23. Moreover, according to Yahoo Finance, Kroger's reputation surged in those times for their COVID-19 response and racial equity support.

Finally, TIF is the ticker for the luxury jewelry company Tiffany \& Co., which admittedly seems an odd pick considering the time-frame of our training data. However, there are two reasons that may explain this, and both of them are completely unrelated to the pandemic. First, the plans of French multinational LVMH (sometimes informally called Louis Vuitton) to acquire Tiffany - plans which were approved by Tiffany's stockholders in February. Second, Tiffany significantly improved their internal cost structure over the preceding three years, making the company more profitable.

\section{Discussion}

The aim of this paper is to analyze different portfolio optimization models and propose a faster, more robust model with a comparable ratio of portfolio expected return to risk (as measured by standard deviation and maximum drawdown). We hope these findings will be helpful in further research on this topic.

To that end, we reviewed the classical Markowitz model together with two well-known variations, while providing some additional analysis on those QP models e.g. confirmed Strong Duality holds by verifying Slater's conditions, proved that $L_1$ regularization has no effect in portfolio optimization (in the case short-selling is not allowed), and devised a heuristic to more objectively choose the $\lambda$ parameter when simultaneously optimizing mean and variance.

Then we introduced a fourth model \citep{Konno1991} that is one of the first examples of an LP formulation of portfolio optimization, and discussed its similarities and (theoretical) advantages compared to the classical QP formulations. This model relies on the so-called `epigraph trick' to turn a non-linear objective (due to the presence of an absolute value) into an LP.

We followed this discussion by presenting a new model that provides an alternative linearization of the classical QP formulation. The model is first presented as a maximin problem (hence non-linear), and then it is turned into an LP by again using the epigraph trick. We observe a solution speed-up by nearly a factor of 10 compared to the classical QPs, while providing similar (if not better) in-sample and out-of-sample performance. The advantages of using a downside risk measure in the model were briefly discussed.

We then proposed a variation of this new model, which turns the LP into a MILP by restricting minimum allocations to 5\% if any budget is allocated to a given stock. In theory, this formulation has worse optimization properties as it should be much harder to solve (MIPs are in general NP-complete). However, due to the particular structure of the problem (including choosing the best possible 'big-M' value) and the use of a highly optimized MIP solver, in practice this variation is about 25 times faster than the LP formulation, and thus 200 times faster than the QPs, while providing a level of performance on par with the other models on most metrics. One exception is in terms of robustness: the MILP formulation seems to be much more robust, at least based on how much the optimal solution changes when the parameter values change. We argue this is due to the minimum allocation restriction, which helps reduce overfitting. The other models, on the other hand, are shown to be quite sensitive to perturbations in parameter values, specifically to changes in the covariance matrix.

The MILP formulation being the most convincing one, we analyzed its optimal solution in depth, arguing why its top three stock choices seem reasonable during COVID-19 times.

A natural extension to this project is considering a multi-period model using Machine Learning and/or stochastic programming. In a multi-period model, the goal is to optimize the expected return and standard deviation of a portfolio considering more than one period. In our case, one possibility is to forecast to some extent the returns of the stocks under consideration for some future periods (possibly with the help of Machine Learning techniques), and optimize the current portfolio accordingly. Another, related possibility is instead of trying to predict the future with sophisticated machinery (pun intended), 'just' make (educated) assumptions on the distribution of returns, and then make use of stochastic programming to capture the inherent non-deterministic nature of the future parameter values (scenario modeling). It is unclear, though, how much there is to be gained from increasing the sophistication of the optimization approach by several orders of magnitude in either of the two ways we have just outlined. But it is clearly interesting and potentially useful to find out.

Finally, the results shown by the MILP formulation in terms of increased speed and robustness while providing a comparable (if not better) in-sample ratio risk-reward seem promising, but further research is necessary to confirm them. A first step in that direction would be to repeat, multiple times, the entire empirical analysis conducted in this project (in sample and out of sample) for different stocks (also different number of stocks) and times (both in horizon length and economic cycle).

\bigskip

\bibliographystyle{plainnat}
\bibliography{Optim_paper}

\end{document}